# Feasibility Analysis of Low Cost Graphical Processing Units for Electromagnetic Field Simulations by Finite Difference Time Domain Method

A. V. Choudhari
Asst. Prof, YCCE,
Hingna Road, Wanadongri,
Nagpur, India.

N. A. Pande
Asst. Prof, YCCE,
Hingna Road, Wanadongri,
Nagpur, India.

M. R. Gupta
M-Tech Student, YCCE,
Hingna Road, Wanadongri,
Nagpur, India.

## ABSTRACT
Among several techniques available for solving Computational Electromagnetics (CEM) problems, the Finite Difference Time Domain (FDTD) method is one of the best suited approaches when a parallelized hardware platform is used. In this paper we investigate the feasibility of implementing the FDTD method using the NVIDIA® GT 520, a low cost Graphical Processing Unit (GPU), for solving the differential form of Maxwell's equation in time domain. Initially a generalized benchmarking problem of bandwidth test and another benchmarking problem of 'matrix left division is discussed for understanding the correlation between the problem size and the performance on the CPU and the GPU respectively. This is further followed by the discussion of the FDTD method, again implemented on both, the CPU and the GT520 GPU. For both of the above comparisons, the CPU used is Intel ® E5300, a low cost dual core CPU.

## General Terms
Computational Electromagnetics, Finite Difference Time Domain method, Algorithm Parallelization.

## Keywords
CUDA, Maxwell's Equation Numerical Solution, GPGPU Computing.

## 1. INTRODUCTION
The clock frequency of modern processors having reached a ceiling, the processor hardware advancement is now oriented towards increasing the number of processing cores. Central Processing Unit (CPU) with up to eight cores is available within costs that are reasonably low. Also, there are Graphical Processing Units with as many as ninety six CUDA® cores within a similar price range. The number of cores however, is very obviously insufficient to measure the processor's ability.

The main bottleneck in any real life problems is usually not just the processor speed/cores, rather it is also the memory bandwidth. That is, the decisive parameter is not just how fast the processor can process the data, what also needs to be considered is the speed with which the data can be transferred to and fro from the processor to the memory. With this variety of options, there arises a need for making proper processor choice best suited for any concerned scientific computations. This leads us to making use of the standard benchmarking problems for such analysis purpose.

## 2. OBSERVATIONS FROM BENCHMARKING OF THE CPU/GPU WITH BANDWITH TEST AND MATRIX LEFT DIVISION

```
Device 0: GeForce GT 520
Quick Mode
Host to Device Bandwidth, 1 Device(s), Paged memory
 Transfer Size (Bytes)        Bandwidth(MB/s)
 33554432                     1493.1

Device to Host Bandwidth, 1 Device(s), Paged memory
 Transfer Size (Bytes)        Bandwidth(MB/s)
 33554432                     1368.1

Device to Device Bandwidth, 1 Device(s)
 Transfer Size (Bytes)        Bandwidth(MB/s)
 33554432                     7841.7
```

**Figure 1: Command prompt snapshot of running a bandwidth test on the GPU GT520 and CPU E5300**

In Figure 1 the 'device' refers to the GPU and the 'host' refers to the CPU. It is seen that the device to device memory transfer speed is little more than 7.5 Gbps, whereas the host to device and device to host memory is merely around 1.4 Gbps. This means that memory transfer, or loading-unloading of the data to and from the GPU is around 5-6 times slower as compared to the memory transfer within the GPU itself

Now, consider a matrix left division operation which is used to solve a system of linear equations (A*x=b). This is system of linear equations is solved with the '/' operator in MATLAB®, by coding as x=A\b. The remaining part of this section deals with benchmarking the GPU GT520 versus CPU E5300 for matrix dimension taken as 1024x1024, 2048x2048, 3072x3072 and 4096x4096 respectively.

In Figure 2 it is clear that the GPU GT520 outperforms the CPU E5300 in the single precision matrix left division benchmarking problems. Hear, one Gigaflops refers to $10^9$ Floating Point Operations/Second. As the size of the problem increases, we see that during the ascent, the slope of the GPU plot is higher than the slope of the CPU plot. That is, an increase in the problem size (matrix size) increases the Gigaflops performance nonlinearly, thus demonstrating that larger data sets are better processed on GPU owing to higher scope of parallelizability. This is however not true for all cases, as it can be seen in Figure 3 where the CPU outperforms the GPU initially in double precision matrix left division problem.

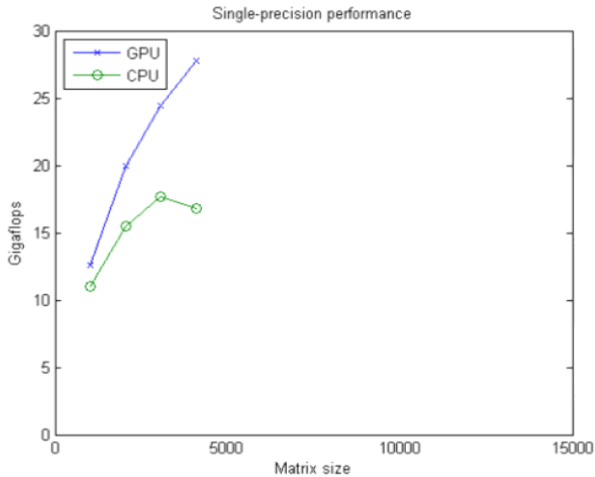

**Figure 2: Gigaflops performance of the GPU GT520 and CPU E5300 for single precision floating point values in matrix left division problem.**

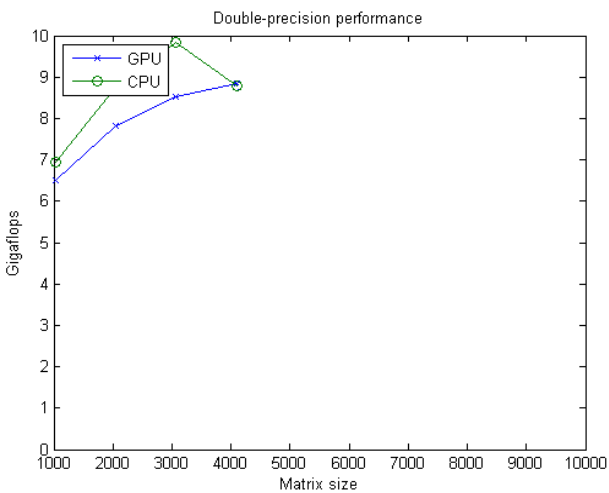

**Figure 3: Gigaflops performance of the GPU GT520 and CPU E5300 for double precision floating point values in matrix left division problem.**

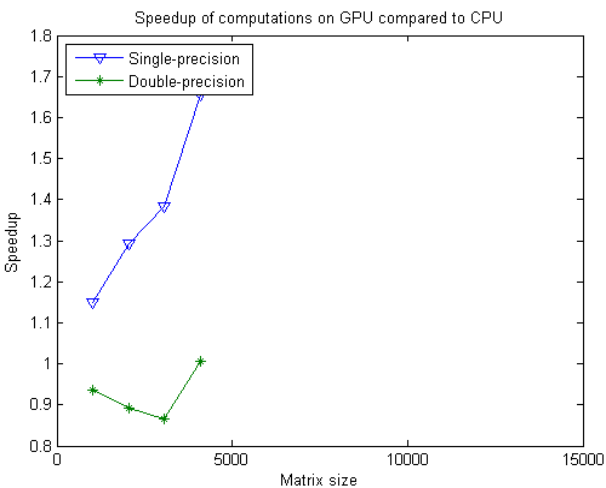

**Figure 4: Speedup achieved on the GPU GT520 versus the CPU E5300 for both single and double precision values.**

In Figure 3, only for the matrix size of 4096x4096, we see that the performance of the GPU is almost same as that of the CPU. This equality is however a mere coincidence and has no engineering significance. For all other smaller problem sizes, the CPU outperforms the GPU. Similarity in performance by the GPU and the GPU for 4096x4096 matrix size is a mere coincidence, where the descend of CPU graph for coincides with the ascend of the GPU graph and is seemingly of no particular engineering importance.

The speedup plotted in Figure 4 is the conclusion of the data from Figure 2 and Figure 3, where speedup is defined as the ratio of the Gigaflops on GPU with the Gigaflops on CPU. Matrix left division benchmarking was not performed for size beyond 4096x4096 because of the memory limitation.

## 3. MATHEMATICAL BACKGROUND OF THE FDTD METHOD AS APPLICABLE IN ELECTROMAGNETIC FIELD SIMULATION PROBLEMS

### 3.1 Central difference approximation

The heart of the FDTD method lies in taking the central difference approximation to the differential form of Maxwell's curl equations [4].

$$\frac{\partial H}{\partial t} = -\frac{1}{\mu}[curl(E) + \sigma * H]$$

$$\frac{\partial E}{\partial t} = \frac{1}{\varepsilon}[curl(H) - \sigma E]$$

Equation 1: Diffrential form of Maxwell's curl equation.

Where **E** is in volt/meter (electric field vector), **H** is in ampere/meter (magnetic field vector,) μ is in henry/meter (magnetic permeability), ε is in farad/meter (electric permittivity), σ is in siemen/meter (electric conductivity).

These Maxwell's equations are further expanded by taking the central difference approximation [3] which is derived from the Taylor Series expansion as given in Equation 2.

$$\frac{df}{dx}(x_0) = \frac{f(x_{0+\frac{\Delta}{2}}) - f(x_{0+\frac{\Delta}{2}})}{\Delta} + O(\Delta^2)$$

Equation 2: The central difference approximation formula.

This leads to the discretization of space. In other words, we now have discrete values of **E** and **H** fields, which can be alternately arranged as proposed by the Yee algorithm. This is shown in Figure 5. The alternate arrangement of **E** and **H** at every grid point or unit cell is separated by half space unit. For a three dimension problem, each grid point will hold six unknown variables, the x, y, and the z component for both **E** and **H** fields. These six equations are shown in Equation 3.

It may be noted hear that each cell in space thus has three **E** components along the x, y, z direction each and three **H** components also along the x, y, z direction.

$$\frac{\partial H_x}{\partial t} = \frac{1}{\mu}(\frac{\partial E_y}{\partial z} - \frac{\partial E_z}{\partial y} - \sigma * H_x)$$

$$\frac{\partial H_y}{\partial t} = \frac{1}{\mu}(\frac{\partial E_z}{\partial x} - \frac{\partial E_x}{\partial z} - \sigma * H_y)$$

$$\frac{\partial H_z}{\partial t} = \frac{1}{\mu}(\frac{\partial E_x}{\partial y} - \frac{\partial E_y}{\partial x} - \sigma * H_z)$$

$$\frac{\partial E_x}{\partial t} = \frac{1}{\varepsilon}(\frac{\partial H_z}{\partial y} - \frac{\partial H_y}{\partial z} - \sigma E_x)$$

$$\frac{\partial E_y}{\partial t} = \frac{1}{\varepsilon}(\frac{\partial H_x}{\partial z} - \frac{\partial H_z}{\partial x} - \sigma E_y)$$

$$\frac{\partial E_z}{\partial t} = \frac{1}{\varepsilon}(\frac{\partial H_y}{\partial y} - \frac{\partial H_x}{\partial y} - \sigma E_z)$$

Equation 3: Six scalar equations containing the vector component of the Maxwell's curl equations.

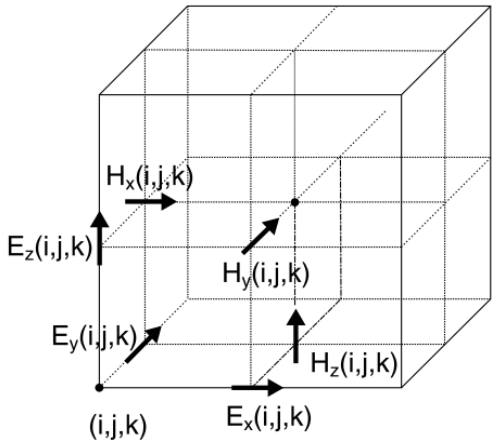

**Figure 5: The Yee Cell.**

### 3.2 The Leapfrog Algorithm

To evaluate the six unknown variables at each grid point, the Leap frog algorithm is used, which is nothing but three simple steps. We begin with using the value of **E** for the grid point containing the electromagnetic source, say a sinusoidal source. The corresponding **H** components can be calculated form Equation 2. Let this grid point containing the source be called Node (0,0,0) where the suffix (0,0,0) is the value along (x,y,z) direction. Once the values of **H** component in Node(0,0,0) is known, the **E** components for the six adjacent nodes (that is Node (1,0,0), Node (-1,0,0), Node (0,1,0), Node (0,-1,0), Node (0,0,1), Node (0,0,-1)) can be calculated again using Equation 2. Each of these **E** field values are then used to calculate the corresponding **H** components values at each of the six adjacent nodes mentioned above.

Thus, we advance in a cyclic fashion and the entire grid is solved for **E** and **H** values. This is nothing but the leap frog method. Programmatically, hear we use update equations where the presently calculated values of **E** and **H** are updated to obtain the future uncalculated values of **E** and **H** respectively.

## 4. IMPLEMENTATION OF THE FDTD METHOD FOR ELECTROMAGNETIC FIELD SIMULATION ON THE CPU

We describe a simple Matlab® program for implementing a one dimensional FDTD simulation. Hear, it is proposed to simulate an electromagnetic wave having the E component in z-direction, the H component along the y-direction and the wave propagation along the x-direction. We initialize the Ez and Hy as an array of zeros, their length being equal to xdim (the dimension along the x axis). This is coded as below:
```
Ez=zeros(1,xdim);
Hy=zeros(1,xdim);
```
We define the electromagnetic source in the center of the x-direction and excite it with a sinusoidal wave as coded below:
```
for n=1:1:time_tot

tstart=1; N_lambda=20;
Ez(xdim/2)=sin(((2*pi*(1/N_lambda)*(n-tstart)*deltat)));
end
```
The update equations for solving the **E** and **H** field are coded as below:
```
for i=1:1:xdim-1
Hy(i)=Hy(i)+(deltat/(delta*mu(i)))*(Ez(i+1)-Ez(i));
end

for i=2:1:xdim
Ez(i)=Ez(i)+(deltat/(delta*epsilon(i)))*(Hy(i)-Hy(i-1));
end
```

Just like the Ez and Hy wes defines, similarly the mu and epsilon are to be defined as an array of corresponding values of µ in henry/meter (magnetic permeability) and ε in farad/meter (electric permittivity) respectively. The output of this program is shown in Figure 6.

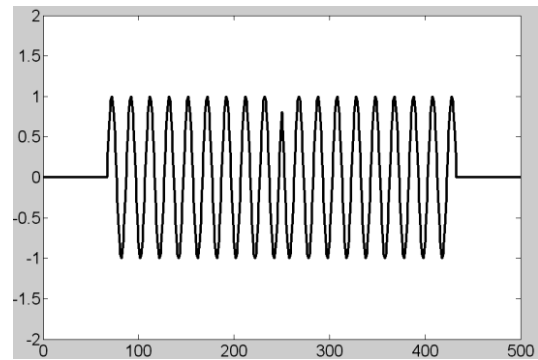

**Figure 6: Output of the one dimension FDTD simulation.**

## 5. IMPLEMENTATION OF THE FDTD METHOD FOR ELECTROMAGNETIC FIELD SIMULATION ON THE GPU

For a GPU implementation of the above program, a simple modification is to be made in the source code. Each array defined earlier is to be redefined as a GPU array as coded below:
```
myEz=gpuArray(zeros(1,xdim));
myHy=gpuArray(zeros(1,xdim));
```
The rest of the code is also implemented with similar modifications.
```
for n=1:1:time_tot
```

```
tstart=1;
N_lambda=20;
myEz(xsource)=sin(((2*pi*(1/N_lambda)*(n-
tstart)*deltat)));
end

 %Update loop for Hy field
for i=1:1:xdim-1
myHy(i)=myHy(i)+(deltat/(delta*mymu(i)))*
(myEz(i+1)-myEz(i));
end
 %Update loop for Ez field
for i=2:1:xdim
myEz(i)=myEz(i)+(deltat/(delta*myepsilon(
i)))*(myHy(i)-myHy(i-1));
end
```

The only difference in the above code is the Matlab® function gpuArray() from the Parallel Computing Toolbox which forms an array on the installed NVIDIA® CUDA® enabled GPU. The output for this program is the same as obtained for the CPU implementation, except for the time of execution being different for each case.

## 6. CONCLUSION

The preliminary results on implementation of one dimensional FDTD on GPU along with the observations of the benchmarking strongly propose that the approach is well suited for implementation of larger real life problems. As observed, although the GPU GT520 speedup is approximately around 1.5 times, however a suitable program implementing the use of both CPU and GPU will lead to a considerable advantage in terms of reduced execution time.

## 7. ACKNOWLEDGMENTS

The authors acknowledge sincere thanks to the Department of Electronics and Telecommunication, Yeshwantrao Chavan College of Engineering, Meghe group of Institutions, Nagpur for providing the research facilities and Dr. P. L. Zade, Head of the Department for his continued support and encouragement for research in the field of electromagnetics and RF antenna design.